# Analysis of the general uncertainty propagation and calculation of the systematic uncertainty of the Neumann Equation of State for surface free energy determination

Jonathan M. Schuster[a,b,c], Carlos E. Schvezov[a], Mario R. Rosenberger[a,b]

[a]*Instituto de Materiales de Misiones (IMAM), Universidad Nacional de Misiones (UNaM), Consejo Nacional de Investigaciones Científicas y Técnicas (CONICET), Félix de Azara 1552, C.P. 3300, Posadas – Misiones – Argentina.*

[b]*Universidad Nacional de Misiones (UNaM), Facultad de Ciencias Exactas, Químicas y Naturales (FCEQyN), Programa de Materiales, Modelización y Metrología, Ruta 12, km 7.5, C.P. 3300, Miguel Lanús, Posadas – Misiones – Argentina.*

[c]*Universidad Nacional del Alto Uruguay (UNAU), Departamento de Ciencias de la Salud, Cátedra de Física Biomédica, Avenida Tejeda 1042, C.P. 3364, San Vicente - Misiones – Argentina.*

*Corresponding author: jschuster@fceqyn.unam.edu.ar

**Keywords**: Surface free energy, Uncertainty analysis, Contact angle, Experimental uncertainty

## Abstract

The determination of the surface free energy ($\gamma_{SV}$) of solid materials provides information to explain and understand a wide variety of phenomena in surface science. One of the most widely used methods to determine the value of $\gamma_{SV}$ is the Neumann equation of state (EQS), whose application requires the values of the contact angle ($\theta$) and the surface tension of a probe liquid ($\gamma_{LV}$). Both parameters are determined experimentally, and their values may be subject to considerable uncertainty. In the present work, the uncertainties of the EQS are analyzed by means of the Taylor series method, using a polynomial adjustment to the EQS previously reported. This analysis allowed us to determine the effects of the $\theta$ and $\gamma_{LV}$ uncertainties on $\gamma_{SV}$ values. The $\gamma_{SV}$ of polyoxymethylene (POM) and polytetrafluoroethylene (PTFE) was calculated using the EQS from four probe liquids and taking into account the propagation of the $\theta$ and $\gamma_{LV}$ uncertainties. Generally, the comparison of the $\gamma_{SV}$ values calculated from the different probe fluids revealed significant differences. As a solution to this inconsistency in the EQS method, we proposed taking into account a systematic standard uncertainty associated with the method equal to 1.3 mJ/m$^2$. This allowed the differences between the $\gamma_{SV}$ values for the different probe liquids to be non-significant and therefore the method to be consistent.

## 1. Introduction

To understand and explain a wide range of phenomena in surface science and engineering, it is very important to know the value of the surface free energy of solid materials ($\gamma_{SV}$) involved in such phenomena [1–5]. In general, the $\gamma_{SV}$ value is not directly measured and therefore must be estimated by means of different approaches [6]. However, due to their simplicity, the methods based on the determination of the contact angle ($\theta$) are the most used [6,7]. These methods are based on Young's equation [7–9]:

$$\gamma_{LV} \cos \theta = \gamma_{SV} - \gamma_{SL} \qquad (1)$$

which relates the liquid-vapor surface tension ($\gamma_{LV}$), solid-vapor surface tension, i.e., surface free energy of the solid ($\gamma_{SV}$), and solid-liquid interfacial tension ($\gamma_{SL}$) with $\theta$. When the solid surface is considered ideal (i.e. homogeneous, smooth, rigid, isotropic, and insoluble and not reactive with the liquid), the contact angle is called Young's angle or equilibrium angle [10].

Among the methods that use $\theta$ values to estimate surface free energy, one of the most widely used is the Neumann equation of state (EQS) [1,7,8,11–13]. This method allows estimating $\gamma_{SV}$ from the determination of the $\theta$ of a single probe liquid, using an EQS that relates $\theta$ to $\gamma_{SV}$ and $\gamma_{LV}$. This method is based on the empirical evidence derived from plotting $\gamma_{LV} \times cos\,\theta$ versus $\gamma_{LV}$ for a given solid, where the points are observed to fit a smooth curve; changing the solid surface gives points that are also consistent with smooth curves and generally parallel to curves from other solid surfaces [13,14]. This behavior suggests that the behavior of $\gamma_{LV} \times cos\,\theta$ is a function of $\gamma_{SV}$ and $\gamma_{LV}$:

$$\gamma_{LV} \times \cos\theta = f(\gamma_{SV}, \gamma_{LV}) \qquad (2)$$

Combining equation 2 with equation 1, we obtain:

$$\gamma_{SL} = F(\gamma_{SV}, \gamma_{LV}) \qquad (3)$$

which suggests the existence of an EQS for the interfacial tensions [14]. By proposing a particular expression for $F(\gamma_{SV}, \gamma_{LV})$ and combining it with Young's equation (equation 1), a relationship between $\theta$ and $\gamma_{SV}$ and $\gamma_{LV}$ is obtained. Three versions of the Neumann EQS, which are differentiated by the arrangement of the physical variables involved in $F(\gamma_{SV}, \gamma_{LV})$ [11,15]. The first of them (N-I), proposed in 1974, is given by the following equation [16]:

$$\frac{(0.015\,\gamma_{SV} - 2.00)\sqrt{\gamma_{LV}\,\gamma_{SV}} + \gamma_{LV}}{\gamma_{LV}(0.015\sqrt{\gamma_{LV}\,\gamma_{SV}} - 1)} - \cos\theta = 0 \qquad (4)$$

The second version (N-II), proposed in 1990, is given by equation [17]:

$$2\sqrt{\gamma_{SV}/\gamma_{LV}}\;e^{-\beta_1(\gamma_{LV} - \gamma_{SV})^2} - 1 - \cos\theta = 0 \qquad (5)$$

While the third version (N-III), proposed in 1999, is given by the following equation [18,19]:

$$2\sqrt{\gamma_{SV}/\gamma_{LV}}\,(1 - \beta_2(\gamma_{LV} - \gamma_{SV})^2) - 1 - \cos\theta = 0 \qquad (6)$$

In equations 4, 5, and 6, $\theta$ is the experimental contact angle between the probe liquid and the tested solid, while $\beta_1$ and $\beta_2$ are constants whose reported values are 0.0001247 $(m^2/mJ)^2$ and 0.0001057 $(m^2/mJ)^2$ respectively [19]. Since the three versions of the Neumann EQS pose non-explicit functions for $\gamma_{SV}$, to find the value of the surface free energy, the equations must be solved by numerical methods.

In general, for the same probe liquid and $\theta$ value, the three versions of the Neumann EQS provide similar values of $\gamma_{SV}$ [18,19]. Previous studies analyzing probe liquids with $\gamma_{LV}$ values from 25 to 72.8 mJ/m$^2$ and $\theta$ values from 1º to 130º reported that the absolute discrepancies between the $\gamma_{SV}$ values obtained by the different EQS do not exceed the value of $\pm 1.3$ $mJ/m^2$ [20]. Regarding the relative discrepancy (quotient of the difference between the values of $\gamma_{SV}$ obtained by the different equations and the value of $\gamma_{SV}$ obtained by one of the equations), higher values were observed when $\theta$ takes values greater than ~100º, that is, in solids with low surface energy; however, the values never exceed $\pm 17\%$ [20]. In both absolute and relative discrepancy, the difference is greater between equations N-III and N-I, intermediate between equations N-II and N-I, and smaller between equations N-III and N-II [20]. In recent literature, the most commonly used EQS are N-II and N-III [7–9,11,12,21]. It is worth mentioning that the usefulness of the EQS to correctly estimate the surface energy in solids has been questioned [21–24].

The $\theta$ values can be determined by different experimental techniques, but one of the most used is the sessile drop technique, where a drop of the order of microliters is deposited on the surface under study [25–27]. The determination of the $\theta$ values by means of this technique is associated with a non-negligible uncertainty, generally taking values from 1º to 2º [28]. However, it is not uncommon for the uncertainty to take higher values of up to 5º [21,29].

In addition to the uncertainty in the $\theta$ value, there is an uncertainty associated with the value of $\gamma_{LV}$ of the probe liquid used. The values of $\gamma_{LV}$ can be determined experimentally using different techniques (e.g., pendant drop method), with an uncertainty that is generally around $\pm 1\ mJ/m^2$ [30–32]. However, in the vast majority of the cases, the $\gamma_{LV}$ of probe liquids are taken from the bibliography. Table 1 shows the $\gamma_{LV}$ values of different probe liquids obtained from different literature sources, where a significant dispersion of values can be observed. This shows that the degree of uncertainty when taking the $\gamma_{LV}$ values from the literature is non-negligible.

***Table 1. Surface tension values of different probe liquids, taken from the literature.***

| Ref. | Surface Tension [mJ/m²] | | | | | | |
|---|---|---|---|---|---|---|---|
| | Wa | Fo | Et | Mi | Gl | Br | Dm |
| [33] | 72.5 | 58.4 | 47.4 | 48.8 | 64.7 | - | 43.7 |
| [34] | 72.4 | 57.3 | 44.7 | 50.8 | - | - | 43.5 |
| [35] | - | 58.2 | - | 50.8 | 63.4 | 44.6 | - |
| [36] | 72.1 | 56.9 | - | 50 | 62.7 | 44.4 | - |
| [37] | 72.8 | 58.2 | - | 50.8 | 63.4 | 44.6 | - |
| [38] | - | 58.4 | 48.2 | - | 63.3 | - | - |
| [39] | 72.8 | 58 | 48 | 50.8 | 64 | 44.4 | 44 |
| [30] | 71.36 | - | - | 48.03 | - | - | - |
| [19] | 72.7 | 59.08 | 47.55 | 49.98 | 65.02 | 44.31 | 42.68 |
| [40] | 71.6 | - | - | - | - | - | 41.7 |
| | | | | | | | |
| Min | 71.36 | 56.90 | 44.70 | 48.03 | 62.70 | 44.31 | 41.70 |
| Max | 72.80 | 59.08 | 48.20 | 50.80 | 65.02 | 44.60 | 44.00 |
| R | 1.44 | 2.18 | 3.50 | 2.77 | 2.32 | 0.29 | 2.30 |

Wa: Water, Fo: Formamide, Et: Ethylene glycol, Mi: methylene iodide, Gl: Glycerol, Br: 1-Bromonaphthalene, Dm: Dimethyl sulfoxide.
Min: Minimum value, Max: Maximum value, R: Range=(Max-Min)

Taking into account the above facts, it becomes necessary to carry out a general analysis of uncertainty in the third Neumann EQS (N-III), which, in the present case, was performed using the Taylor series method (TSM) [41,42]. This allows determining and analyzing the effect of the experimental uncertainties of the input variables ($\theta$ and $\gamma_{LV}$) on the final result ($\gamma_{SV}$). To this end, we used a polynomial fit of the N-III EQS $\left( PM\gamma_{SV}(\theta, \gamma_{LV}) \right)$ reported in the literature [20], which is explicit for $\gamma_{SV}$ and mathematically derivable, which enables the use of TSM. The general uncertainty analysis allows examining the relative standard uncertainty in $\gamma_{SV}$ as a function of the uncertainties in $\theta$ and $\gamma_{LV}$, the uncertainty magnification factors for $\theta$ and $\gamma_{LV}$, and the uncertainty percentage contribution of $\theta$ and $\gamma_{LV}$ in $\gamma_{SV}$ [41]. In addition, we analyzed the standard and expanded uncertainty in the values of $\gamma_{SV}$ calculated due to the uncertainties in both $\theta$ and $\gamma_{LV}$ for two polymeric surfaces, one of polyoxymethylene (POM) and the other of polytetrafluoroethylene (PTFE), determined from the measurement of the $\theta$ values of four different probe liquids. Considering that when using different probe liquids, the N-III method should provide equal values of $\gamma_{SV}$ for the same surface, we evaluated whether the difference between the values of $\gamma_{SV}$ obtained for the four liquids is statistically significant (taking the uncertainties in $\theta$ and $\gamma_{LV}$). Finally, we analyze and propose the need to consider a systematic standard uncertainty inherent to the N-III method.

## 2. Materials and methods

### 2.1. $PM\gamma_{SV}(\theta, \gamma_{LV})$ polynomial and general uncertainty analysis

To carry out the general uncertainty analysis of the Neumann EQS and calculate the values of $\gamma_{SV}$, we used a polynomial model fitted to the N-III EQS $\left( PM\gamma_{SV}(\theta, \gamma_{LV}) \right)$, given by the following equations [20]:

$$PM\gamma_{SV}(\theta, \gamma_{LV}) = A(\gamma_{LV}) \times \theta^5 + B(\gamma_{LV}) \times \theta^4 + C(\gamma_{LV}) \times \theta^3 + D(\gamma_{LV}) \times \theta^2 +$$

$$E(\gamma_{LV}) \times \theta^1 + F(\gamma_{LV}) \tag{7}$$

where the coefficients A, B, C, D, E, and F of this polynomial function are in turn functions of $\gamma_{LV}$:

$$A(\gamma_{LV}) = -3.6735 \times 10^{-14}\gamma_{LV}{}^3 + 5.4452 \times 10^{-12}\gamma_{LV}{}^2 - 1.9749 \times 10^{-10}\gamma_{LV} +$$

$$1.5627 \times 10^{-9} \tag{7.1}$$

$$B(\gamma_{LV}) = +1.1040 \times 10^{-11}\gamma_{LV}{}^3 - 1.6918 \times 10^{-9}\gamma_{LV}{}^2 + 6.1290 \times 10^{-8}\gamma_{LV} -$$

$$5.3196 \times 10^{-7} \tag{7.2}$$

$$C(\gamma_{LV}) = -1.0814 \times 10^{-9}\gamma_{LV}{}^3 + 1.6463 \times 10^{-7}\gamma_{LV}{}^2 - 5.0646 \times 10^{-6}\gamma_{LV} +$$

$$5.6396 \times 10^{-5} \tag{7.3}$$

$$D(\gamma_{LV}) = +4.2138 \times 10^{-8}\gamma_{LV}{}^3 - 5.6112 \times 10^{-6}\gamma_{LV}{}^2 + 2.8255 \times 10^{-5}\gamma_{LV} -$$

$$2.1287 \times 10^{-3} \tag{7.4}$$

$$E(\gamma_{LV}) = -5.7967 \times 10^{-7}\gamma_{LV}{}^3 + 7.4232 \times 10^{-5}\gamma_{LV}{}^2 - 2.5385 \times 10^{-3}\gamma_{LV} +$$

$$3.1449 \times 10^{-2} \tag{7.5}$$

$$F(\gamma_{LV}) = +9.9952 \times 10^{-1}\gamma_{LV} - 9.3025 \times 10^{-3} \tag{7.6}$$

where $\theta$ is given in degrees (deg) and $\gamma_{LV}$ in $mJ/m^2$. In the equations of $A(\gamma_{LV})$, $B(\gamma_{LV})$, $C(\gamma_{LV})$, $D(\gamma_{LV})$, $E(\gamma_{LV})$, and $F(\gamma_{LV})$, the units of the independent term are $mJ/m^2$, while the units of the coefficients that multiply $\gamma_{LV}{}^3$, $\gamma_{LV}{}^2$, and $\gamma_{LV}$ are $m^2/(mJ^2\ deg^n)$, $m/(mJ\ deg^n)$, and $1/deg^n$ respectively, where n equals 5, 4, 3, 2, 1 and 0 for A, B, C, D, E, and F respectively. The domain of $PM\gamma_{SV}(\theta, \gamma_{LV})$ is $1° \leq \theta \leq 130°$ and $25\ mJ/m^2 \leq \gamma_{LV} \leq 72.8\ mJ/m^2$ [20] and the systematic standard uncertainty [43] associated with the polynomial fit ($u_{PM}$) is 0.01257 mJ/m² [20].

The first step for the general uncertainty analysis consists in obtaining an expression that allows calculating the combined standard uncertainty of the surface free energy determined by the N-III method $\left(u_{\gamma_{SV}}\right)$. This is achieved by considering the function $PM\gamma_{SV}(\theta, \gamma_{LV})$ and using the TSM [41–44]:

$$\left(u_{\gamma_{SV}}\right)^2 = \left(\frac{\partial PM\gamma_{SV}}{\partial \theta}\right)^2 (u_\theta)^2 + \left(\frac{\partial PM\gamma_{SV}}{\partial \gamma_{LV}}\right)^2 \left(u_{\gamma_{LV}}\right)^2 \tag{8}$$

where $u_\theta$ is the standard uncertainty of $\theta$ and $u_{\gamma_{LV}}$ is the standard uncertainty of $\gamma_{LV}$ of the probe liquid. In equation 8, it is assumed that the input variables $\theta$ and $\gamma_{LV}$ are independent, i.e. not correlated [41,45]. In addition, equation 8 does not include an additional term for the systematic standard uncertainty corresponding to the polynomial function (fitting standard uncertainty), which is calculated as the $u_{PM}$ raised to square, resulting equal to $0.000158 \; mJ^2/m^4$ and being negligible for the uncertainty value [20]. The general analysis of uncertainties was carried out in the entire domain of the function $PM\gamma_{SV}(\theta, \gamma_{LV})$.

## 2.1.1. Analysis of the relative standard uncertainty in $\gamma_{SV}$

If both terms of equation 8 are divided by the square of the function $PM\gamma_{SV}$, while the first term on the right-hand side is multiplied by $(\theta/\theta)^2$ and the second term is multiplied by $(\gamma_{LV}/\gamma_{LV})^2$, the following expression is obtained [41,46,47]:

$$\left(\frac{u_{\gamma_{SV}}}{PM\gamma_{SV}}\right)^2 = \left(\frac{\theta}{PM\gamma_{SV}} \frac{\partial PM\gamma_{SV}}{\partial \theta}\right)^2 \left(\frac{u_\theta}{\theta}\right)^2 + \left(\frac{\gamma_{LV}}{PM\gamma_{SV}} \frac{\partial PM\gamma_{SV}}{\partial \gamma_{LV}}\right)^2 \left(\frac{u_{\gamma_{LV}}}{\gamma_{LV}}\right)^2 \tag{9}$$

where $\frac{u_{\gamma_{SV}}}{PM\gamma_{SV}} = \varepsilon_{u_{\gamma_{SV}}}$ is the relative standard uncertainty in the surface free energy, $u_\theta/\theta = \varepsilon_{u_\theta}$ is the relative standard uncertainty in the contact angle and $u_{\gamma_{LV}}/\gamma_{LV} = \varepsilon_{u_{\gamma_{LV}}}$ is the relative standard uncertainty in the surface tension of the probe liquid. This equation allows analyzing, in a general way, the behavior of the relative standard uncertainty in the surface free energy as a function of the relative uncertainty in the input parameters $\theta$ and $\gamma_{LV}$.

## 2.1.2. Uncertainty magnification factors

The factors in parentheses that multiply $(u_\theta/\theta)$ and $(u_{\gamma_{LV}}/\gamma_{LV})$ in equation 9 are called uncertainty magnification factors (UMFs) and, in this case, are defined as [41]:

$$UMF_\theta = \frac{\theta}{PM\gamma_{SV}} \frac{\partial PM\gamma_{SV}}{\partial\theta} \tag{10}$$

$$UMF_{\gamma_{LV}} = \frac{\gamma_{LV}}{PM\gamma_{SV}} \frac{\partial PM\gamma_{SV}}{\partial\gamma_{LV}} \tag{11}$$

where $UMF_\theta$ and $UMF_{\gamma_{LV}}$ are the uncertainty magnification factors for $\theta$ and $\gamma_{LV}$, respectively. An absolute value of $UMF_i$ greater than 1 indicates that the influence of the relative uncertainty of the variable $i$ is magnified when it is propagated in the equation of interest, while an absolute value lower than 1 indicates that the influence of the variable decreases as it propagates in the equation of interest [41].

## 2.1.3. Uncertainty percentage contribution

Another way to analyze the effect of the uncertainties of the input variables is through the uncertainty percentage contribution (UPC) factor [41]. This factor can be obtained by dividing equation 8 by $\left(u_{\gamma_{SV}}\right)^2$, which gives:

$$1 = \frac{\left(\frac{\partial PM\gamma_{SV}}{\partial\theta}\right)^2 (u_\theta)^2 + \left(\frac{\partial PM\gamma_{SV}}{\partial\gamma_{LV}}\right)^2 \left(u_{\gamma_{LV}}\right)^2}{\left(u_{\gamma_{SV}}\right)^2} \tag{12}$$

or by dividing equation 9 by the relative uncertainty in $\gamma_{SV}$ squared $\left(\varepsilon_{u_{\gamma_{SV}}}{}^2\right)$, which gives:

$$1 = \frac{\left(\frac{\theta}{PM\gamma_{SV}} \frac{\partial PM\gamma_{SV}}{\partial\theta}\right)^2 \left(\varepsilon_{u_\theta}\right)^2 + \left(\frac{\gamma_{LV}}{PM\gamma_{SV}} \frac{\partial PM\gamma_{SV}}{\partial\gamma_{LV}}\right)^2 \left(\varepsilon_{u_{\gamma_{LV}}}\right)^2}{\left(\varepsilon_{u_{\gamma_{SV}}}\right)^2} \tag{13}$$

Equations 12 and 13 allow defining the UPC of the variables $\theta$ and $\gamma_{LV}$ as:

$$UPC_\theta = \frac{\left(\frac{\partial PM\gamma_{SV}}{\partial\theta}\right)^2 (u_\theta)^2}{\left(u_{\gamma_{SV}}\right)^2} \times 100 = \frac{\left(\frac{\theta}{PM\gamma_{SV}} \frac{\partial PM\gamma_{SV}}{\partial\theta}\right)^2 \left(\varepsilon_{u_\theta}\right)^2}{\left(\varepsilon_{u_{\gamma_{SV}}}\right)^2} \times 100 \tag{14}$$

$$UPC_{\gamma_{LV}} = \frac{\left(\frac{\partial PM_{\gamma_{SV}}}{\partial \gamma_{LV}}\right)^2 \left(u_{\gamma_{LV}}\right)^2}{\left(u_{\gamma_{SV}}\right)^2} \times 100 = \frac{\left(\frac{\gamma_{LV}}{PM_{\gamma_{SV}}} \frac{\partial PM_{\gamma_{SV}}}{\partial \gamma_{LV}}\right)^2 \left(\varepsilon_{u_{\gamma_{LV}}}\right)^2}{\left(\varepsilon_{u_{\gamma_{SV}}}\right)^2} \times 100 \qquad (15)$$

The factors $UPC_{\theta}$ and $UPC_{\gamma_{LV}}$ account for the percentage contribution of the uncertainty in $\theta$ and $\gamma_{LV}$ in the squared uncertainty in the surface free energy, so $UPC_{\theta} + UPC_{\gamma_{LV}} = 100\ \%$ [41].

## 2.2. Experimental determination of the surface free energy in PTFE and POM

### 2.2.1. Sample preparation

The solid surface of PTFE was polished using SiC papers with granulometry ranging from #320 to #1500 and then with 1 µm diamond paste lubricated with ethylene glycol for 20 minutes at 60 rpm. The solid surface of POM was polished using SiC papers with granulometry ranging from #500 to #1500 and then placed between two glass slides and kept in a furnace at 180 ℃ for 10 minutes, to obtain a good smooth surface. The final roughness of the samples was determined by profilometry. Five measurements were made on each sample, obtaining an average roughness (Ra) of 123 nm (standard deviation of 40 nm) for PTFE and a Ra of 269 nm (standard deviation of 39 nm) for POM. Prior to the deposition of a drop to measure $\theta$, the PTFE and POM surfaces were cleaned with detergent and water, rinsed with ethanol, and dried in hot air, then cleaned with acetone and dried again with hot air [48].

### 2.2.2. Contact angle measurement and surface free energy determination

The $\theta$ values on the PTFE and POM surfaces were measured by the sessile drop technique, using a laboratory-made and calibrated goniometer [49]. Static contact angles were obtained, that is to say that, during the measurement, the contact line of the three phases of the sessile drop (solid-vapor-liquid) remained fixed with respect to the surface of the solid [50]. The probe liquids used were water (Wa, deionized), formamide (Fo, >99.5%, Biopack), methylene iodine (Mi, >99%, Sigma-Aldrich) and 1-bromonaphthalene (Br, >95%, Merck). Drops of 2 µL deposited with a micropipette were used. The contact angle was measured between 5 and 10 seconds after the drop was deposited on the surface. On each surface, 12 drops were deposited and measured for each probe liquid, and the $\theta$ value reported is the arithmetic mean of those measurements. From the $\theta$ values determined and using the function $PM\gamma_{SV}(\theta, \gamma_{LV})$ (equation 7), the values of $\gamma_{SV}$ were obtained for each probe liquid and surface evaluated. Images of the sessile droplets and θ values measured for this work are available in a data repository [51].

### 2.2.3. Determination of the standard uncertainty in the contact angle ($u_\theta$), in the surface tension of the probe liquids ($u_{\gamma_{LV}}$), and in the surface free energy ($u_{\gamma_{SV}}$)

Since the $\theta$ value is the result of several measurements, the value of its standard uncertainty ($u_\theta$) is calculated as the quotient between the standard deviation (*SD*) and the positive square root of the number of drops measured (*n*) [41,43]:

$$u_\theta = SD/\sqrt{n} \tag{14}$$

The standard uncertainty $u_{\gamma_{LV}}$ is not obtained from repeated measurements but determined *a priori* from a distribution [41,43]. Normal distributions with 99% probability were assumed, where the value of $\gamma_{LV}$ is included in the interval where $a_-$ and $a_+$ are the lower and upper bounds, respectively. The limits $a_-$ and $a_+$ were obtained by subtracting and adding 2 $mJ/m^2$, respectively, to the values of $\gamma_{LV}$ reported by Kwok and Neumann [19]: water (72.7 $mJ/m^2$), formamide (59.08 $mJ/m^2$), methylene iodine (49.98 $mJ/m^2$) and 1-bromonaphthalene (44.31 $mJ/m^2$). The value of $\pm 2$ $mJ/m^2$ was chosen considering the variability observed in the $\gamma_{LV}$ data reported in the bibliography (see Table 1) [19,30,33–40]. The standard uncertainty is obtained from the following equation [43]:

$$u_{\gamma_{LV}} = (a_+ - a_-)/(2 \times 2.58) \tag{15}$$

Therefore, the value of $u_{\gamma_{LV}}$ is equal to 0.775 $mJ/m^2$ and is the same for all the probe liquids used.

The standard uncertainty in the surface free energy $(u_{\gamma_{SV}})$ is a combined standard uncertainty [43] since it arises from the combination of the uncertainties in the contact angle $(u_\theta)$ and in the surface tension of the probe liquid used $(u_{\gamma_{LV}})$. The value of $u_{\gamma_{SV}}$ was determined from equation 8.

### 2.2.4. Determination of the expanded uncertainty in the contact angle $(U_\theta)$ and in the surface free energy $(U_{\gamma_{SV}})$

The expanded uncertainty $(U_q)$ defines an interval around the result of a measurement $q$ in which a significant fraction of the distribution of values that could reasonably be attributed to the measurand is expected to be found [41,43]. The fraction can be thought of as the confidence level of the interval. The value of $(U_q)$ is calculated by multiplying the standard uncertainty associated with the measurand $q$ $(u_q)$ by a coverage factor $k_q$:

$$U_q = k_q\, u_q \tag{16}$$

This defines an interval given by the maximum $(U_{qmax})$ and minimum $(U_{qmin})$ extremes:

$$U_{qmax} = \bar{q} + U_q \tag{17}$$

$$U_{qmin} = \bar{q} - U_q \tag{18}$$

where $\bar{q}$ is generally the arithmetic mean of a series of measurements or the value determined from other quantities by means of a functional relationship. This interval $\bar{q} \pm U_q$ has a confidence level $p$ given by the coverage factor used [43]. In this work, the expanded uncertainties were determined with a confidence level of 99 %.

In the case of $\theta$, the value of $k_q$ is equal to the value $t(\nu)$ of the t distribution with $\nu$ degrees of freedom, which defines an interval from $-t(\nu)$ to $+t(\nu)$ that comprises a fraction of 99 % of distribution. The number of degrees of freedom associated with the determination of $\theta$ $(\nu_\theta)$ is $n - 1 = 11$, so the value of the coverage factor $k_\theta$ is equal to 3.11 for all the probe liquids used in this work [43].

In the case of the surface free energy, its calculation is associated with a combined standard uncertainty $(u_{\gamma_{SV}})$ obtained through equation 8. So, the effective degrees of freedom must be calculated $\left(\nu_{\gamma_{SV}}^{eff}\right)$ to obtain the value of $k_{\gamma_{SV}}$ in each particular case by using the Welch-Satterthwaite approximation [41,43]. However, in order to simplify calculations and taking into account the Central Limit Theorem, it is possible to assume an approximately normal distribution for the variable $\gamma_{SV}$ with a mathematical expectation $E(\gamma_{SV}) = PM\gamma_{SV}(\theta, \gamma_{LV})$ and a variance $\left(u_{\gamma_{SV}}\right)^2$ given by equation 8 [43]. So, a value of $k_{\gamma_{SV}}$ equal to 2.58 can be used to obtain the uncertainty expanded to 99 % of the surface free energy $(U_{\gamma_{SV}})$. In this work, this same assumption was made in all cases where it was necessary to determine the coverage factor associated with a combined standard uncertainty.

### 2.2.5. Comparison of the $\gamma_{SV}$ values obtained for different probe liquids on POM and PTFE surfaces

The surface energy values obtained using different probe liquids were compared by calculating the difference between the values obtained using $PM\gamma_{SV}(\theta, \gamma_{LV})$ (equation 7). To determine whether the difference between two values of $\gamma_{SV}$ was significant, we calculated the standard uncertainty of such difference; if the interval built with the expanded uncertainty of the difference includes zero, it is concluded that the difference is not significant [41]. Two different methods (A and B) were used to calculate the standard uncertainty of the difference between the values of $\gamma_{SV}$.

### 2.2.5.1. Method A

For this method, we followed the recommendation of Coleman & Steele [41], which consists in generating a new function from the difference between the values to be compared. For the case in which the surface energy values obtained by liquids $i$ and $j$ are compared, a function that results from the difference between the function $PM\gamma_{SV}(\theta, \gamma_{LV})$ evaluated with liquid $i$ $\left(PM\gamma_{SV}^i(\theta_i, \gamma_{LV}^i)\right)$ and with the liquid $\left(PM\gamma_{SV}^j(\theta_j, \gamma^j{}_{LV})\right)$ is constructed:

$$r_{i-j}(\theta_i, \gamma_{LV}^i, \theta_j, \gamma_{LV}^j) = PM\gamma_{SV}^i(\theta_i, \gamma^i{}_{LV}) - PM\gamma_{SV}^j(\theta_j, \gamma^j{}_{LV}) \tag{19}$$

In this way, the combined standard uncertainty of $r_{i-j}$ will be given by the following equation:

$$u_{r_{i-j}} =$$

$$\sqrt{\left(\frac{\partial PM\gamma_{SV}^i}{\partial \theta_i}\right)^2 (u_{\theta_i})^2 + \left(\frac{\partial PM\gamma_{SV}^i}{\partial \gamma^i{}_{LV}}\right)^2 \left(u_{\gamma^i{}_{LV}}\right)^2 + \left(\frac{\partial PM\gamma_{SV}^j}{\partial \theta_j}\right)^2 (u_{\theta_j})^2 + \left(\frac{\partial PM\gamma_{SV}^j}{\partial \gamma^j{}_{LV}}\right)^2 \left(u_{\gamma^j{}_{LV}}\right)^2}$$

$$\tag{20}$$

### 2.2.5.2. Method B

For this method, we simply considered the difference of the surface energy values $\gamma_{SV}^i$ and $\gamma_{SV}^j$ calculated with liquids $i$ and $j$ by means of the polynomial $PM\gamma_{SV}(\theta, \gamma_{LV})$:

$$s_{i-j}(\gamma^i{}_{SV}, \gamma_{SV}^j) = \gamma^i{}_{SV} - \gamma_{SV}^j \tag{21}$$

where $s_{i-j}(\gamma^i{}_{SV}, \gamma_{SV}^j)$ will have the same numerical value as $r_{i-j}(\theta_i, \gamma_{LV}^i, \theta_j, \gamma_{LV}^j)$. However, the combined standard uncertainty of $s_{i-j}$ is calculated in a different way (from other parameters) and given by the following equation:

$$u_{s_{i-j}} = \sqrt{\left(u_{\gamma_{SV}^i}\right)^2 + \left(u_{\gamma_{SV}^j}\right)^2} \tag{22}$$

For both methods, the coverage factor used ($k_r$ or $k_s$) was 2.58 and was used to obtain the expanded uncertainty (99%) of the difference between the surface free energy values obtained by two different liquids on the same solid ($U_r$ o $U_s$).

## 3. Results and discussion

### 3.1. General uncertainty analysis

#### 3.1.1. Behavior of the relative uncertainty associated with the determination of the surface free energy

Figure 1 shows the percentage relative standard uncertainty in the surface free energy ($\varepsilon_{u_{\gamma_{SV}}}$ %) calculated using equation 7 in a contour plot, considering that the relative uncertainties in $\theta$ and $\gamma_{LV}$ are of 1%. As a reference, curves where the surface energy of the solid shows constant values (green = 10 $mJ/m^2$, light blue = 25 $mJ/m^2$, and red = 50 $mJ/m^2$) are shown. It can be seen that $\varepsilon_{u_{\gamma_{SV}}}$ takes values greater than 1 % in the entire range analyzed, except for two regions located at low angles (1° to ~20°), where $25\ mJ/m^2 < \gamma_{LV} < \sim28\ mJ/m^2$ and $\sim55\ mJ/m^2 < \gamma_{LV} < \sim73\ mJ/m^2$, respectively. The value of $\varepsilon_{u_{\gamma_{SV}}}$ increases with the value of $\theta$ and is practically independent of the value of $\gamma_{LV}$ of the probe liquid. For a given value of $\gamma_{SV}$, the values of $\varepsilon_{u_{\gamma_{SV}}}$ decrease when using probe liquids with a lower value of $\gamma_{LV}$.

The maximum values for $\varepsilon_{u_{\gamma_{SV}}}$ were obtained for surfaces with low surface free energy and probe liquids with high surface tension. For example, for the case of $\gamma_{SV} \cong 10\ mJ/m^2$ (see green line in Figure 1), the standard relative uncertainty was of ~7 % when using water as a probe liquid ($\gamma_{LV} = 72.8\ mJ/m^2$), and decreases to ~3.5 % when using a probe liquid with a surface tension of $\sim40\ mJ/m^2$.

To analyze what happens when the relative uncertainty in $\theta$ and $\gamma_{LV}$ increases (always considering that $\varepsilon_{u_{\gamma_{LV}}} = \varepsilon_{u_\theta}$), a value of $\varepsilon_{u_{\gamma_{SV}}}$ can simply be multiplied by the percentage observed on the contour plot in Figure 1 by the times the relative uncertainty in $\theta$ and $\gamma_{LV}$ increased. For example, if it is considered that the relative uncertainty in $\theta$ and $\gamma_{LV}$ doubles from 1 % to 2 %, we will have a value of $\varepsilon_{u_{\gamma_{SV}}}$ of ~14 % when using water as probe liquid and a value of ~7% when using a probe liquid with $\gamma_{LV} \cong 40 \; mJ/m^2$.

In addition, using Figure 1 it is possible to analyze the behavior of the expanded relative uncertainty $\left( \varepsilon_{U_{\gamma_{SV}}} \right)$ for the particular case in which $\varepsilon_{u_{\gamma_{LV}}} = \varepsilon_{u_\theta} = 1$ %. This analysis is done by multiplying the values observed in the contour plot by the coverage factor k, which corresponds to the confidence level of the expanded uncertainty $U$ considered.

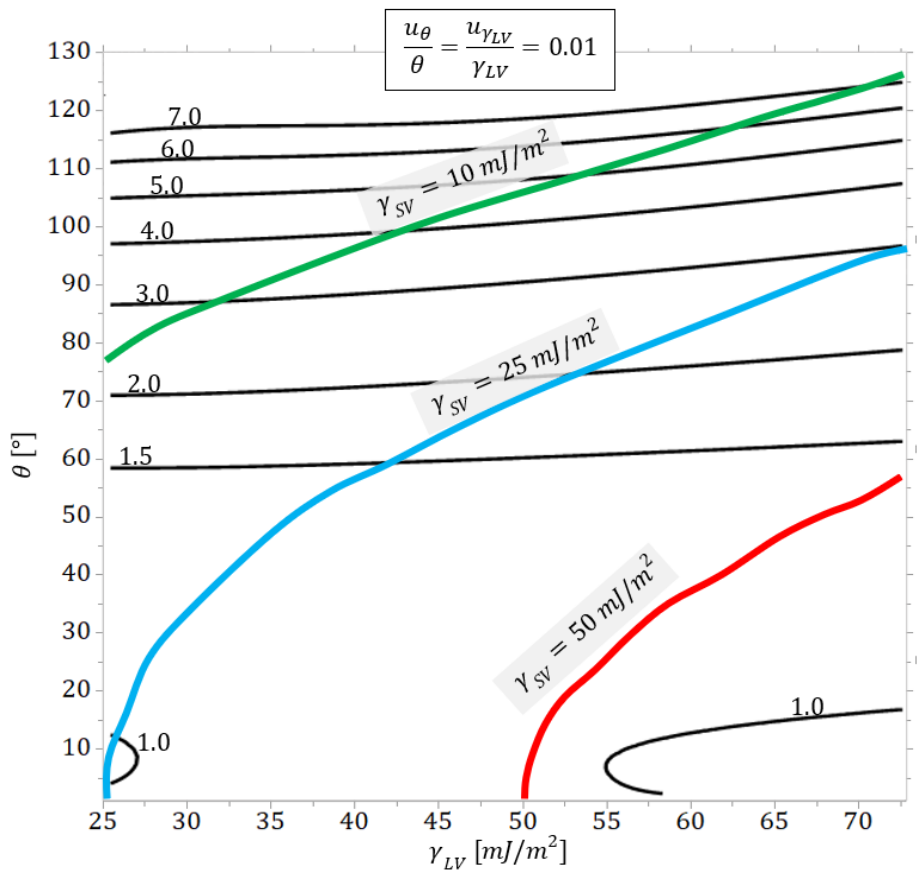

Figure 1. Contour plot of the percentage relative standard uncertainty (solid black line) of the surface free energy $\left( \varepsilon_{u_{\gamma_{SV}}} = \left( u_{\gamma_{SV}}/PM\gamma_{SV} \right) \times 100 \; \% \right)$ as a function of $\theta$ and $\gamma_{LV}$. The plot was built for relative uncertainties of $\theta$ and $\gamma_{LV}$ equal to 1 % each. The *colored lines indicate a given value of $\gamma_{SV}$ (green: $\gamma_{SV}=10$ mJ/$m^2$, light blue: $\gamma_{SV}=25$ mJ/$m^2$, red: $\gamma_{SV}=50$ mJ/$m^2$).*

### 3.1.2. Uncertainty amplification factor

Figures 2 and 3 show the contour plot of the absolute value of the uncertainty amplification factor for the contact angle ($UMF_\theta$) and for the surface tension of the liquid ($UMF_{\gamma_{LV}}$), respectively. In both cases, the behavior of the UMF is shown as a function of the $\theta$ and $\gamma_{LV}$ values.

In Figure 2, the value of $|UMF_\theta|$ increases with the value of $\theta$ and is practically independent of the value of $\gamma_{LV}$. For values of $\theta$ lower than ~60°, the value of $|UMF_\theta|$ is lower than 1, so that, in that region, the relative uncertainty in $\theta$ is damped when propagated in the function $PM\gamma_{SV}(\theta, \gamma_{LV})$. Due to this behavior, if a surface with a determined surface free energy is considered, it is possible to choose a particular probe liquid to decrease the value of $UMF_\theta$ and thus decrease the value of the uncertainty in the value of $\gamma_{SV}$, achieving more accurate values. For example, if, for a surface with a value of $\gamma_{SV} = \sim25$ mJ/$m^2$, we work with a probe liquid whose $\gamma_{LV}$ is $\sim40$ mJ/$m^2$, the value of $|UMF_\theta|$ will be lower than 1, while if we use water ($\gamma_{LV} = 72.8$ mJ/$m^2$), the value of $|UMF_\theta|$ will be ~2.5. For surfaces with high surface free energy ($\gamma_{SV} > 45$ mJ/$m^2$), regardless of the probe liquid chosen, the value of $|UMF_\theta|$ will be lower than 1. However, the smaller the value of $\gamma_{LV}$, the closer to zero the value of $|UMF_\theta|$ and therefore the greater the damping of the relative uncertainty in $\theta$, achieving more precise values. For surfaces with low surface energy ($\gamma_{SV} < 15$ mJ/$m^2$), the values of $|UMF_\theta|$ will always be greater than 1 and increase as the value of $\gamma_{LV}$ of the probe liquid increases.

Figure 3 shows that $\left|UMF_{\gamma_{LV}}\right|$ increases with increasing values of $\theta$ and $\gamma_{LV}$. In general, the value of $\left|UMF_{\gamma_{LV}}\right|$ is greater than 1 in the entire zone analyzed, and the regions with values lower than 1 are at low $\theta$ (from 1º to ~20º) with $\gamma\gamma_{LV} < \sim 28\ mJ/m^2$ and $\gamma_{LV} > \sim 55\ mJ/m^2$. For the case of a particular solid and the same as what happens with $\left|UMF_{\theta}\right|$, it is possible to choose a particular probe liquid to decrease the value of $\left|UMF_{\gamma_{LV}}\right|$. The values of $\left|UMF_{\gamma_{LV}}\right|$ are generally smaller for solids with high surface energy and, for the same solid, the value is minimized when using a probe liquid with lower $\gamma_{LV}$ values.

It is noteworthy that, for a given solid, using water as the probe liquid maximizes the values of both $\left|UMF_{\theta}\right|$ and $\left|UMF_{\gamma_{LV}}\right|$, thus giving results that have greater associated uncertainty.

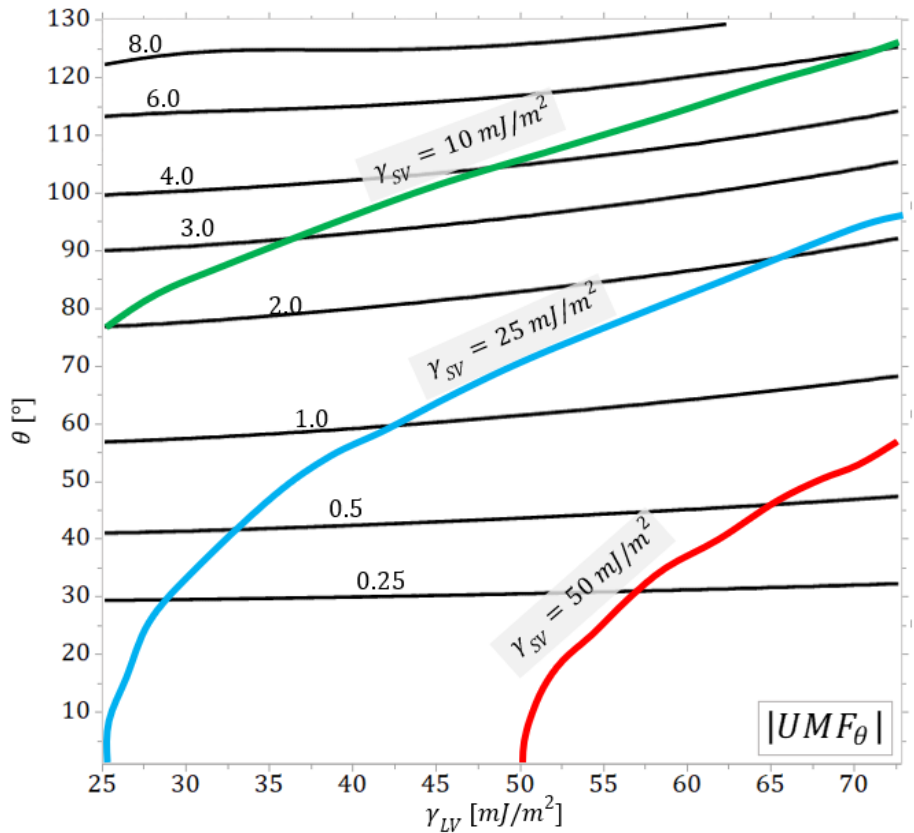

*Figure 2. Contour plot of the absolute value of the contact angle uncertainty amplification factor ($\left|UMF_{\theta}\right|$) as a function of $\theta$ and $\gamma_{LV}$ (solid black line). The colored*

*lines indicate a given value of $\gamma_{SV}$ (green: $\gamma_{SV}=10\ mJ/m^2$, light blue: $\gamma_{SV}=25\ mJ/m^2$ and red: $\gamma_{SV}=50\ mJ/m^2$).*

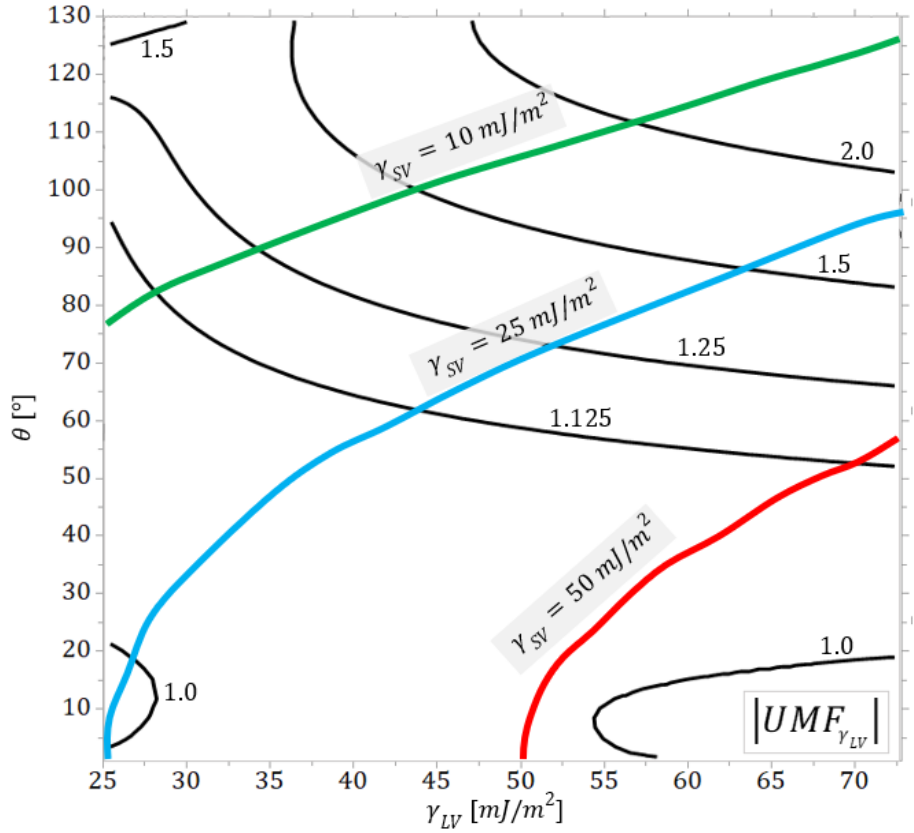

*Figure 3. Contour plot of the absolute value of the probe liquid surface tension uncertainty amplification factor $\left(\left|UMF_{\gamma_{LV}}\right|\right)$ as a function of $\theta$ and $\gamma_{LV}$ (solid black line). The colored lines indicate a given value of $\gamma_{SV}$ (green: $\gamma_{SV}=10\ mJ/m^2$, light blue: $\gamma_{SV}=25\ mJ/m^2$ and red: $\gamma_{SV}=50\ mJ/m^2$).*

### 3.1.3. Uncertainty percentage contribution in surface free energy

Figure 4 shows a contour plot of the UPC of the variable $\theta$ to the surface free energy $(UPC_\theta)$ as a function of $\theta$ and $\gamma_{LV}$. The value of $UPC_\theta$ was calculated according to equation 12 considering the equality of the relative standard uncertainty in $\theta$ and $\gamma_{LV}$. The value of $UPC_\theta$ increases with the value of $\theta$ and is slightly dependent on the value of $\gamma_{LV}$ of the probe liquid, mainly for values of $\theta$ greater than $\sim50°$, where $UPC_\theta$ increases as the value of $\gamma_{LV}$ increases. The comparison of $UPC_\theta$ and $UPC_{\gamma_{LV}}$ shows that, in the region where $\theta > \sim 60°$, $UPC_\theta$ dominates, while, when $\theta < \sim 60°$, $UPC_{\gamma_{LV}}$ prevails.

From the point of view of the surface energy of the solid, on surfaces with low surface energy ($\gamma_{SV} \cong 10 \ mJ/m^2$, see green line in Figure 4), the value of $UPC_\theta$ remains approximately constant as the value of $\gamma_{LV}$ of the probe liquid varies, while for surfaces with medium and high surface energy ($\gamma_{SV} > \sim 25 \ mJ/m^2$, see light blue line in Figure 4), the value of $UPC_\theta$ is considerably reduced as the value of $\gamma_{LV}$ of the probe liquid decreases. On surfaces with high surface energy ($\gamma_{SV} > \sim 50 \ mJ/m^2$, see red line in Figure 4), the contribution of the variable $\gamma_{LV}$ ($UPC_{\gamma_{LV}}$) prevails.

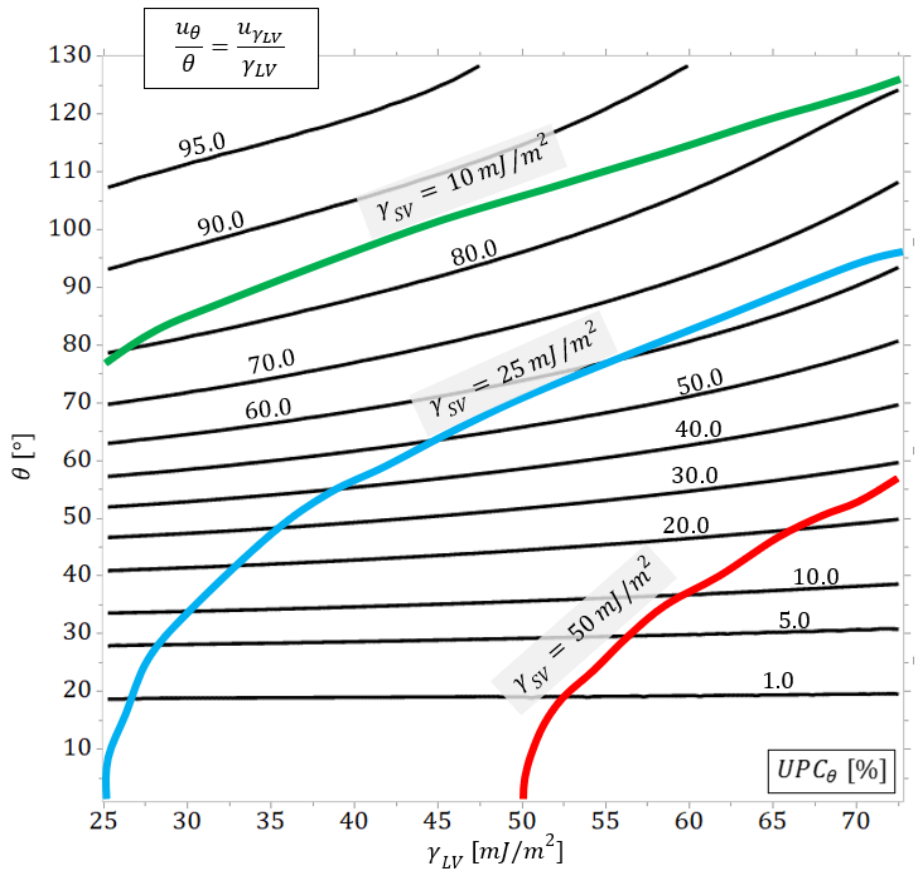

*Figure 4. Contour plot of the uncertainty percentage contribution of the variable $\theta$ to the surface free energy ($UPC_\theta$) as a function of $\theta$ and $\gamma_{LV}$ (solid black line). The plot*

*was constructed for equal relative uncertainties of $\theta$ and $\gamma_{LV}$. The colored lines indicate a given value of $\gamma_{SV}$ (green: $\gamma_{SV}$=10 mJ/m², light blue: $\gamma_{SV}$=25 mJ/m², red: $\gamma_{SV}$=50 mJ/m²). For the analysis of $UPC_{\gamma_{LV}}$, we considered that $UPC_{\gamma_{LV}} = 100\ \% - UPC_{\theta}$.*

## 3.2. Surface free energy of PTFE and POM

### 3.2.1. Contact angles

Table 2 shows the $\theta$ values obtained for PTFE and POM. A clear decrease in the $\theta$ value is observed with the decrease in the surface tension of the probe liquid. The standard uncertainty of $\theta$ for the four probe liquids is on average 0.73º in PTFE and 0.86º in POM. So, it can be considered that the standard uncertainty of $\theta$ is generally found at values close to 1º and is therefore approximately independent of the probe liquid and of the value of $\theta$. The values of the mean expanded uncertainty of $\theta$ were 2.33º in PTFE and 2.67º in POM. The highest values of relative standard uncertainty and relative expanded uncertainty were observed with 1-bromonaphthalene (17% in POM and 4.6% in PTFE), mainly because it provides the lowest $\theta$ values (~18º in POM and ~67º in PTFE).

Figure 5 (A and B) shows the values of $\theta$ obtained in PTFE and POM, respectively, in comparison with the values reported in the bibliography [52–62]. Similar values were observed for each probe liquid. The absolute average difference between the values obtained in this work and those reported in the literature, considering the four liquids, was 4.43º ($SD$=2.86º) for PTFE and 2.44º ($SD$=2.26º) for POM.

Figure 6 shows a plot of $\gamma_{LV} \times cos\ (\theta)$ vs. $\gamma_{LV}$ for the four liquids on the PTFE and POM surfaces. The data were fitted to a second-degree polynomial by using the method of least squares. The smoothness of the fitting curves suggests that $\gamma_{LV} \times cos\ (\theta)$ is a function of $\gamma_{LV}$ and $\gamma_{SV}$, as assumed by the Neumann EQS (see equation 2) [13,14].

**Table 2. Values of contact angles of water (Wa), formamide (Fo), methylene iodine (Mi) and 1-bromonaphthalene (Br) determined on POM and PTFE.**

| Liquid | Parameters | PTFE | POM |
|--------|------------|------|-----|
| Wa | $\bar{x}_\theta$ | 108.30 | 75.98 |
|    | $u_\theta$ | 0.63 | 0.99 |
|    | $\varepsilon_{u_\theta}$ | 0.58 | 1.30 |
|    | $U_\theta$ | 1.95 | 3.07 |
|    | $\varepsilon_{U_\theta}$ | 1.80 | 4.04 |
| Fo | $\bar{x}_\theta$ | 91.19 | 58.08 |
|    | $u_\theta$ | 0.88 | 0.68 |
|    | $\varepsilon_{u_\theta}$ | 0.96 | 1.18 |
|    | $U_\theta$ | 2.73 | 2.13 |
|    | $\varepsilon_{U_\theta}$ | 2.99 | 3.66 |
| Mi | $\bar{x}_\theta$ | 80.92 | 38.88 |
|    | $u_\theta$ | 0.44 | 0.77 |
|    | $\varepsilon_{u_\theta}$ | 0.54 | 1.98 |
|    | $U_\theta$ | 1.35 | 2.39 |
|    | $\varepsilon_{U_\theta}$ | 1.67 | 6.14 |
| Br | $\bar{x}_\theta$ | 67.61 | 18.23 |
|    | $u_\theta$ | 0.99 | 1.00 |
|    | $\varepsilon_{u_\theta}$ | 1.47 | 5.49 |
|    | $U_\theta$ | 3.08 | 3.11 |
|    | $\varepsilon_{U_\theta}$ | 4.56 | 17.06 |

$\bar{x}_\theta$ [°]: arithmetic mean of $\theta$ measurements,

$u_\theta$ [°]: standard uncertainty of $\theta$ (equation 14),

$\varepsilon_{u_\theta} = (u_\theta / \bar{x}_\theta) \times 100$ [%]: relative standard uncertainty of $\theta$,

$U_\theta$ [°]: expanded uncertainty of $\theta$ with a confidence level of 99% (equation 16),

$\varepsilon_{U_\theta} = (U_\theta / \bar{x}_\theta) \times 100$ [%]: relative expanded uncertainty of $\theta$.

The coverage factor $(k_\theta)$ is 3.11 in all cases.

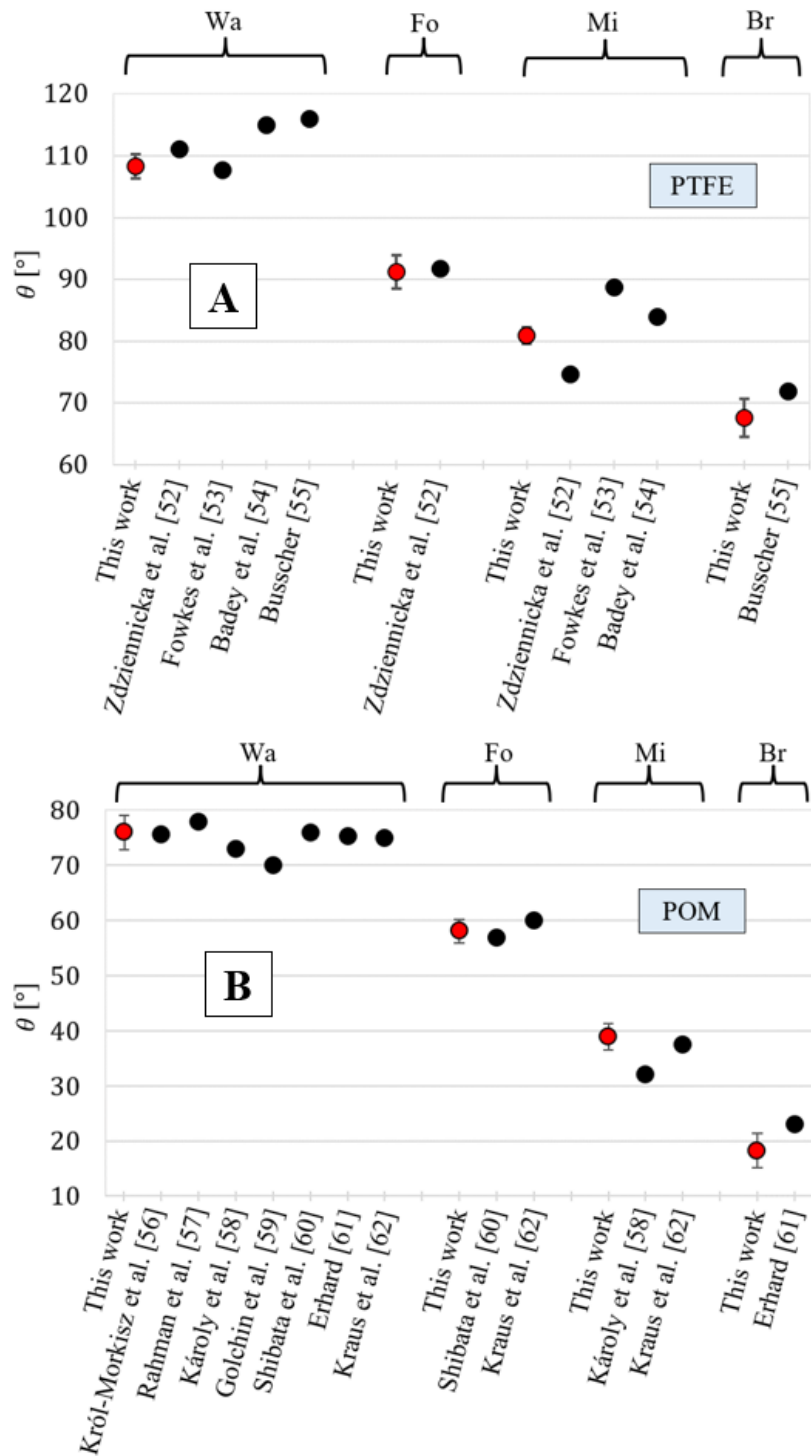

*Figure 5. Comparison of contact angle values on PTFE (A) and POM (B). The values obtained in this work are shown in red (error bar corresponds to uncertainty expanded to 99%), whereas the values taken from the literature are shown in black.*

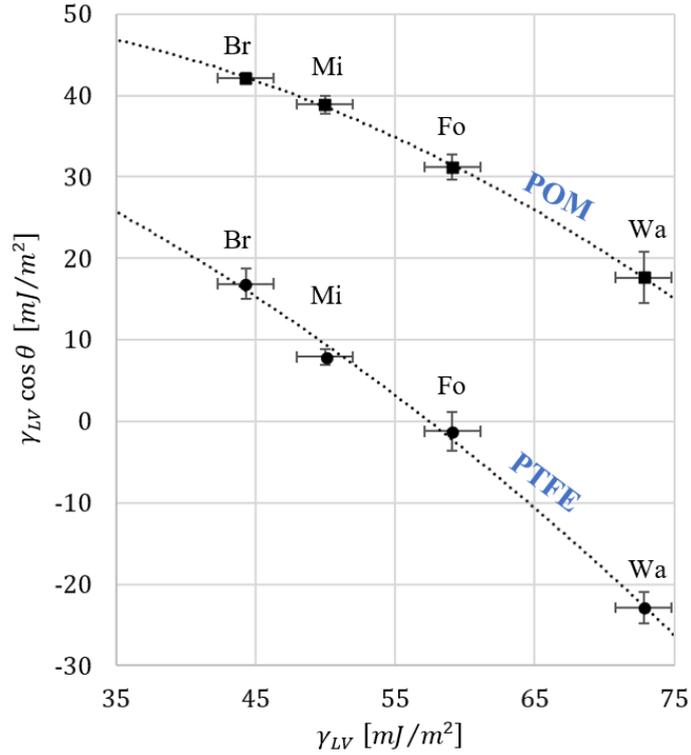

*Figure 6. Plot of $\gamma_{LV}\cos(\theta)$ versus $\gamma_{LV}$ for the four liquids used in this work on POM and PTFE surfaces. The vertical and horizontal error bars correspond to the expanded uncertainty with a confidence level of 99 % in $\gamma_{LV} \times \cos(\theta)$ and in $\gamma_{LV}$, respectively.*

### 3.2.2. Surface free energy

Figure 7 shows the $\gamma_{SV}$ values for PTFE and POM, calculated with the different probe liquids considering the uncertainty in both the $\theta$ and the surface tension values of the probe liquid. The $\gamma_{SV}$ values for PTFE are between 18.23 $mJ/m^2$ and 23.26 $mJ/m^2$ with an average value of 20.49 $mJ/m^2$, while those for POM are between 37.33 $mJ/m^2$ and 42.16 $mJ/m^2$, with an average value of 39.46 $mJ/m^2$. On both surfaces, the values of $\gamma_{SV}$ are shown ordered from the lowest to the highest and correspond to those obtained with water, formamide, methylene iodine and 1-bromonaphthalene, which shows the increase in the value of $\gamma_{SV}$ calculated with the decrease in the solid-liquid interfacial tension of the probe liquid. The values of $\gamma_{SV}$ obtained on both surfaces are in the ranges of values reported in the bibliography, from 18 $mJ/m^2$ to 26 $mJ/m^2$ for PTFE and from 38 $mJ/m^2$ to 45 $mJ/m^2$ for POM [63].

The standard uncertainty in $\gamma_{SV}$ is on average 0.69 $mJ/m^2$ for PTFE and 0.81 $mJ/m^2$ for POM. So, it can be considered that the standard uncertainty of $\gamma_{SV}$ is approximately independent of the probe liquid used for its determination and is generally found at values close to ~1 $mJ/m^2$. The expanded uncertainty of $\gamma_{SV}$ is between 1.56 $mJ/m^2$ and 1.94 $mJ/m^2$ for PTFE and between 1.96 $mJ/m^2$ and 2.30 $mJ/m^2$ for POM. For more details on the values calculated, see Table A.1 in Appendix A.

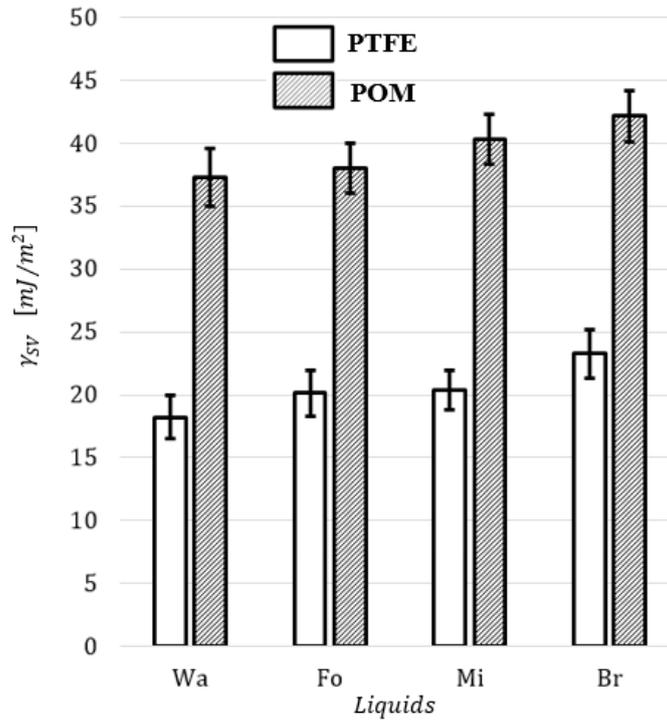

Figure 7. Surface free energy values ($\gamma_{SV}$) for POM and PTFE calculated with the liquids water (Wa), formamide (Fo), methylene iodine (Mi) and 1-bromonaphthalene (Br). The error bars correspond to the expanded uncertainty ($U_{\gamma_{SV}}$) with a confidence level of 99%. For more details, see Table A.1.

### 3.2.3. Comparison of surface free energy calculated from different probe liquids

Six comparisons of the four probe liquids were made for each surface: Wa-Fo, Wa-Mi, Wa-Br, Fo-Mi, Fo-Br and Mi-Br. Figure 8 shows the results of the comparisons with the expanded uncertainties calculated with the methods A $(r_{i-j} \pm U_r)$ and B $(s_{i-j} \pm U_s)$. The values of $r_{i-j}$ and $s_{i-j}$ are identical and take values ranging from -0.68 $mJ/m^2$ to -4.82 $mJ/m^2$ for POM and from -0.25 $mJ/m^2$ to -5.03 $mJ/m^2$ for PTFE. On the other hand, the uncertainty values (both standard and expanded) present discrepancies, with the uncertainty for $s_{i-j}$ being greater than the uncertainty calculated for $r_{i-j}$, with the only exception of the Mi-Br comparison in POM, where the calculated uncertainties are equal. The $U_r/U_s$ (and $u_r/u_s$) ratio in POM takes a mean value of 0.94, a minimum value of 0.91 and a maximum value of 1.00, whereas that in PTFE takes a mean value of 0.85, a minimum value of 0.78 and a maximum value of 0.91, that is, method B $(s_{i-j})$ generally overestimates the uncertainty of the difference in surface free energy values. The values of the comparisons between liquids and their uncertainties are detailed in Table B.1 in Appendix B.

Figure 9 shows a scatter plot of the values of $r_{i-j}$ (and $s_{i-j}$) for both surfaces where a moderate positive correlation is evident (Pearson's coefficient=0.66). The liquid pair that deviates from the linear trend is Fo-Mi because the difference observed in PTFE (0.25 $mJ/m^2$) is very low compared to that observed in POM (2.3 $mJ/m^2$). If the Fo-Mi difference is removed, the Pearson's correlation coefficient increases to a value of 0.79. This approximately linear relationship of the values of $r_{i-j}$ on both surfaces shows that, in general, the pairs of liquids that provide higher (or lower) values of $r_{i-j}$ for POM also do so for PTFE.

In the cases in which the intervals of the expanded uncertainty for comparisons between the different pairs of probe liquids $(r_{i-j} \pm U_r)$ include zero, it can be affirmed that the values of surface energy estimated by the different liquids are not significantly different. As can be seen in Figure 7 (and Table B.1), not all intervals include zero. The comparisons that do not include zero are Wa-Mi, Wa-Br and Fo-Br for both surfaces (PTFE and POM) and Mi-Br only for PTFE. So, in these cases, the difference between the values of surface free energy is significant. These results demonstrate the importance of the correct comparison of the surface free energy values. If the analysis of the comparison between the values of $\gamma_{SV}$ had been limited to comparing and analyzing the overlapping of the confidence intervals $\gamma_{SV} \pm U_{\gamma_{SV}}$ shown in Figure 7 (and Table A.1), we would have erroneously concluded that only the difference between the Wa-Br liquids is significant (both in POM and PTFE) and that the other comparisons give values of $\gamma_{SV}$ that are not significantly different from each other due to the uncertainty associated with each determination.

### 3.2.4. Determination of standard systematic uncertainty in the N-III method

The significant differences observed in Figure 8 may be explained by the fact that not all sources of uncertainty, such as the uncertainty associated with the hypothesis that gives rise to the Neumann EQS [19,23,41,64,65], are taken into account. In view of this, it is possible to consider that the Neumann EQS method has a standard systematic uncertainty $(u_m)$ [41].

To obtain a lower bound for $u_m$, we can calculate the minimum value of $u_m$ such that the differences observed in the values of surface free energy obtained by the different liquids in PTFE and POM are not significant, that is, the expanded uncertainty of the difference includes zero. For this, the uncertainty $u_m$ is included in equation 20 $\left(u_{r_{i-j}}\right)$, in which the values of the uncertainty in the difference in the values of surface energy obtained by liquids $i$ and $j$ are calculated as:

$$u'_{r_{i-j}} =$$

$$\sqrt{\left(\frac{\partial PM\gamma^i_{SV}}{\partial \theta_i}\right)^2 (u_{\theta_i})^2 + \left(\frac{\partial PM\gamma^i_{SV}}{\partial \gamma^i_{LV}}\right)^2 \left(u_{\gamma^i_{LV}}\right)^2 + (u_{m-i})^2 + \left(\frac{\partial PM\gamma^j_{SV}}{\partial \theta_j}\right)^2 (u_{\theta_j})^2 + \left(\frac{\partial PM\gamma^j_{SV}}{\partial \gamma^j_{LV}}\right)^2 \left(u_{\gamma^j_{LV}}\right)^2 + (u_{m-j})^2}$$

$$(23)$$

where $u_{m-i}$ and $u_{m-j}$ are the standard systematic uncertainty associated with the N-III EQS due to the application of the model when liquids $i$ and $j$ are used, respectively. Considering that $u_{m-i} = u_{m-j} = u_m$ and substituting equation 20 in equation 23, we obtain:

$$u'_{r_{i-j}} = \sqrt{\left(u_{r_{i-j}}\right)^2 + 2(u_m)^2} \qquad (24)$$

By multiplying the value of $u'_{r_{i-j}}$ by the coverage factor $k'_r = 2.58$ [41], the expanded uncertainty of $r_{i-j}$ (at 99 %) is obtained considering the uncertainty in the surface tension of the probe liquid, the uncertainty in the angle contact and the systematic uncertainty associated with the N-III method. To obtain an estimate of the minimum value of $u_m$, we propose that the expanded uncertainty of $r_{i-j}$ $(U'_{r_{i-j}})$ be equal to $r_{i-j}$ so that the interval of the expanded uncertainty includes zero:

$$U'_{r_{i-j}} = u'_{r_{i-j}} k'_r = r_{i-j} \qquad (25)$$

By replacing equation 24 in equation 25, we obtain:

$$\sqrt{\left(u_{r_{i-j}}\right)^2 + 2(u_m)^2} \, k'_r = r_{i-j} \qquad (26)$$

By solving equation 26 for $u_m$, we obtain:

$$u_m = \frac{\sqrt{(r_{i-j})^2 - (k_r')^2 (u_{r_{i-j}})^2}}{\sqrt{2}\, k_r'} \qquad (27)$$

Equation 27 can be used to calculate the value of $u_m$ such that equation 25 is fulfilled. The values of $u_m$ calculated in each particular case are shown in Table 3, in which it can be observed that, if a value of $u_m$ is greater than or equal to 1.24 mJ/m², the interval of the expanded uncertainty of all the comparisons made will include zero. Therefore, the N-III EQS method with this consideration and for the experimental values obtained in this work is self-consistent, that is, when using different probe liquids on the same surface, the N-III EQS provides values of $\gamma_{SV}$ that are not significantly different. It should be noted that the values of $u_m$ obtained and shown in Table 3 depend on the standard uncertainties considered for the values of $\theta$ ($u_\theta$) and surface tension of the probe liquids $(u_{\gamma_{LV}})$. The dependence is especially important for the uncertainty $u_{\gamma_{LV}}$ since it is defined *a priori*.

It should also be noted that, although in this work the analysis of the difference between the values of surface free energy ($r_{i-j}$ or $s_{i-j}$) was used to study the self-consistency of the N-III method, it can also be used to analyze whether there are significant differences between surface free energy values between different surfaces or between surfaces subjected to different surface treatments [41].

**Table 3. Minimum systematic uncertainty values of the N-III equation of state.**

| Material | Liquids | $u_m$ $[mJ/m^2]$ |
|---|---|---|
| POM | Wa-Mi | 0.27 |
| | Wa-Br | 1.05 |
| | Fo-Br | 0.85 |
| PTFE | Wa-Mi | 0.30 |
| | Wa-Br | 1.24 |
| | Fo-Br | 0.52 |
| | Mi-Br | 0.49 |
| $u_m$: systematic uncertainty associated with the N-III equation of state (equation 27) | | |

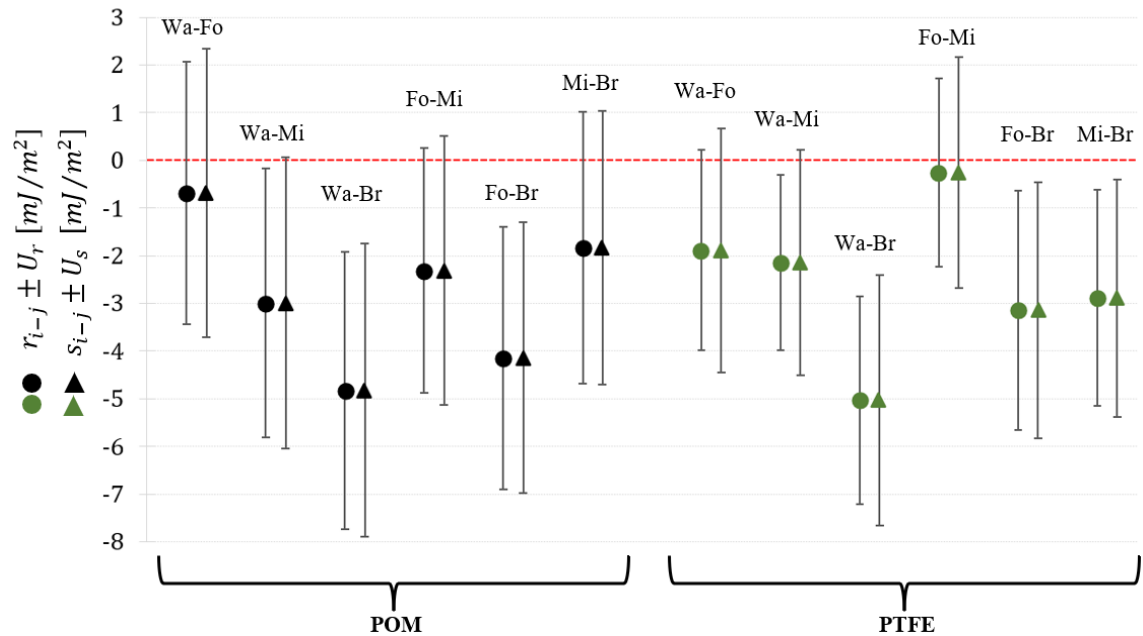

*Figure 8. Difference between the surface free energy values obtained using different probe liquids on the same surface. $r_{i-j}$ (equation 19) and $s_{i-j}$ (equation 21) are represented. The bars correspond to the expanded uncertainty $U_r$ and $U_s$ (99% confidence level).*

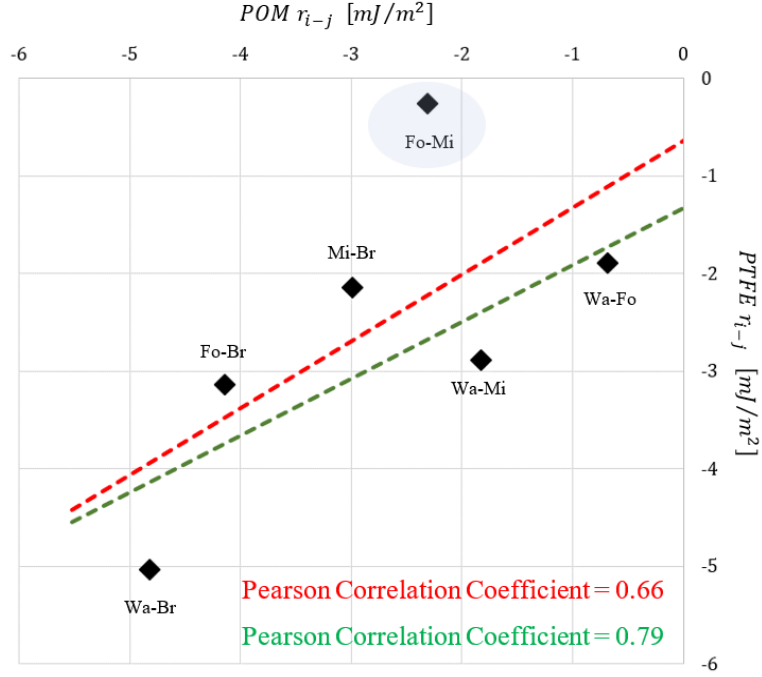

*Figure 9. Scatter plot of the values of $r_{i-j}$ in POM and PTFE. Red dotted line: linear trend considering the six points. Green dotted line: linear trend without considering the value of $r_{Fo-Mi}$.*

### 3.2.5. Recalculation of the standard and expanded uncertainty in $\gamma_{SV}$ considering the systematic error in the Neumann EQS method ($u_m$)

To calculate the standard uncertainty of the surface free energy, it is necessary to include the systematic standard uncertainty of the N-III method. To do this, equation 6 is modified in such a way that it includes the systematic error of the $u_m$ model:

$$\left(u'_{\gamma_{SV}}\right)^2 = \left(\frac{\partial PM\gamma_{SV}}{\partial \theta}\right)^2 (u_\theta)^2 + \left(\frac{\partial PM\gamma_{SV}}{\partial \gamma_{LV}}\right)^2 \left(u_{\gamma_{LV}}\right)^2 + (u_m)^2 \qquad (28)$$

The value considered for $u_m$ was 1.30 $mJ/m^2$, which is higher than the values found for $u_m$ (see Table 3). The coverage factor $k'_{\gamma_{SV}}$ used to obtain the expanded uncertainty values ($U'_{\gamma_{SV}}$) was 2.58.

The average of the values of $u'_{\gamma_{SV}}$ in PTFE was 1.53 mJ/m$^2$, while that in POM was 1.47 mJ/m$^2$. So, it can be considered that the standard uncertainty of $\gamma_{SV}$ (considering $u_{\gamma_{LV}}$, $u_\theta$ and $u_m$) is approximately independent of the probe liquid used for its determination and a value close to ~1.5 mJ/m$^2$ is taken. The values of $u'_{\gamma_{SV}}$ (and $U'_{\gamma_{SV}}$) in general double the values of $u_{\gamma_{SV}}$ (and $U_{\gamma_{SV}}$), and the $u'_{\gamma_{SV}}/u_{\gamma_{SV}}$ (and $U'_{\gamma_{SV}}/U_{\gamma_{SV}}$) ratios take a mean value of 2.16 for PTFE and 1.90 for POM. Likewise, when considering the uncertainty $u_m$, in PTFE the relative standard uncertainty increases on average ~4 percentage points and the relative expanded uncertainty increases on average ~20 percentage points, whereas, in POM the relative standard uncertainty increases on average ~2 percentage points and the relative expanded uncertainty increases on average ~10 percentage points.

To analyze the relative uncertainty in the surface energy considering the uncertainty $u_m$, equation 9 is modified, obtaining:

$$\left(\frac{u'_{\gamma_{SV}}}{PM\gamma_{SV}}\right)^2 = \left(\frac{\theta}{PM\gamma_{SV}}\frac{\partial PM\gamma_{SV}}{\partial\theta}\right)^2\left(\frac{u_\theta}{\theta}\right)^2 + \left(\frac{\gamma_{LV}}{PM\gamma_{SV}}\frac{\partial PM\gamma_{SV}}{\partial\gamma_{LV}}\right)^2\left(\frac{u_{\gamma_{LV}}}{\gamma_{LV}}\right)^2 + \left(\frac{u_m}{PM\gamma_{SV}}\right)^2 \quad (29)$$

This equation allows analyzing, in a general way, the behavior of the relative standard uncertainty in the surface free energy as a function of the relative uncertainty of the input parameters ($\theta$ and $\gamma_{LV}$) and of $u_m$. Figure 10 shows the percentage relative standard uncertainty in the surface free energy ($\varepsilon_{u'_{\gamma_{SV}}}$ %) calculated using equation 29 in a contour plot, considering that the relative uncertainties in $\theta$ and $\gamma_{LV}$ are 1% and that the standard uncertainty $u_m$ is 1.3 $mJ/m^2$. A great difference can be observed in the values with respect to those in Figure 1, where $u_m$ was not considered, which reveals the great influence of $u_m$ on the percentage relative standard uncertainty of the surface free energy. In addition, in Figure 10, the probe liquids are located according to their values of $\gamma_{LV}$ and $\theta$ (on POM and PTFE) used in this work, and it can be observed that the relative standard uncertainty values according to the contour plot are very similar to the calculated values: 3.5 to 4% for POM and 6 to 8% for PTFE (see Table 4). The situation represented in Figure 10 can be considered as the minimum expected relative standard uncertainty in the surface free energy for most of the determinations made with the N-III method.

*Table 4. Surface free energy values and standard and expanded uncertainty, considering a systematic standard uncertainty in the Neumann model $u_m$ of 1.30 $mJ/m^2$.*

| Liquid | Parameters | PTFE | POM |
|---|---|---|---|
| Wa | $\gamma_{SV}$ | 18.23 | 37.33 |
| | $u'_{\gamma_{SV}}$ | 1.46 | 1.58 |
| | $\varepsilon_{u'_{\gamma_{SV}}}$ | 8.01 | 4.22 |
| | $U'_{\gamma_{SV}}$ | 3.77 | 4.07 |
| | $\varepsilon_{U'_{\gamma_{SV}}}$ | 20.67 | 10.90 |
| | $U'_{\gamma_{SV}}/U_{\gamma_{SV}}$ | 2.20 | 1.77 |
| Fo | $\gamma_{SV}$ | 20.11 | 38.02 |
| | $u'_{\gamma_{SV}}$ | 1.49 | 1.51 |
| | $\varepsilon_{u'_{\gamma_{SV}}}$ | 7.39 | 3.96 |
| | $U'_{\gamma_{SV}}$ | 5.32 | 5.39 |
| | $\varepsilon_{U'_{\gamma_{SV}}}$ | 26.45 | 14.19 |
| | $U'_{\gamma_{SV}}/U_{\gamma_{SV}}$ | 2.07 | 1.98 |
| Mi | $\gamma_{SV}$ | 20.37 | 40.33 |
| | $u'_{\gamma_{SV}}$ | 1.43 | 1.52 |
| | $\varepsilon_{u'_{\gamma_{SV}}}$ | 7.04 | 3.76 |
| | $U'_{\gamma_{SV}}$ | 6.57 | 6.94 |
| | $\varepsilon_{U'_{\gamma_{SV}}}$ | 32.24 | 17.22 |
| | $U'_{\gamma_{SV}}/U_{\gamma_{SV}}$ | 2.37 | 1.94 |
| Br | $\gamma_{SV}$ | 23.26 | 42.16 |
| | $u'_{\gamma_{SV}}$ | 1.50 | 1.52 |
| | $\varepsilon_{u'_{\gamma_{SV}}}$ | 6.46 | 3.61 |
| | $U'_{\gamma_{SV}}$ | 8.38 | 8.49 |
| | $\varepsilon_{U'_{\gamma_{SV}}}$ | 36.02 | 20.14 |
| | $U'_{\gamma_{SV}}/U_{\gamma_{SV}}$ | 2.00 | 1.92 |

$\gamma_{SV}$ [$mJ/m^2$]: surface free energy (equation 7).

$u'_{\gamma_{SV}}$ [$mJ/m^2$]: combined standard uncertainty of $\gamma_{SV}$ (equation 28).

$\varepsilon_{u'_{\gamma_{SV}}} = \left( u_{\gamma_{SV}}/\gamma_{SV} \right) \times 100$ [%] : relative standard uncertainty of $\gamma_{SV}$.

$U'_{\gamma_{SV}}$ [$mJ/m^2$]: expanded uncertainty of $\gamma_{SV}$ with a confidence level of 99%.

$\varepsilon_{U'_{\gamma_{SV}}} = \left( U_{\gamma_{SV}}/\gamma_{SV} \right) \times 100$ [%]: relative expanded uncertainty of $\gamma_{SV}$.

Coverage factor (k) is 2.58.

$U_{\gamma_{SV}}$ = expanded uncertainty without considering the $u_m$ uncertainty.

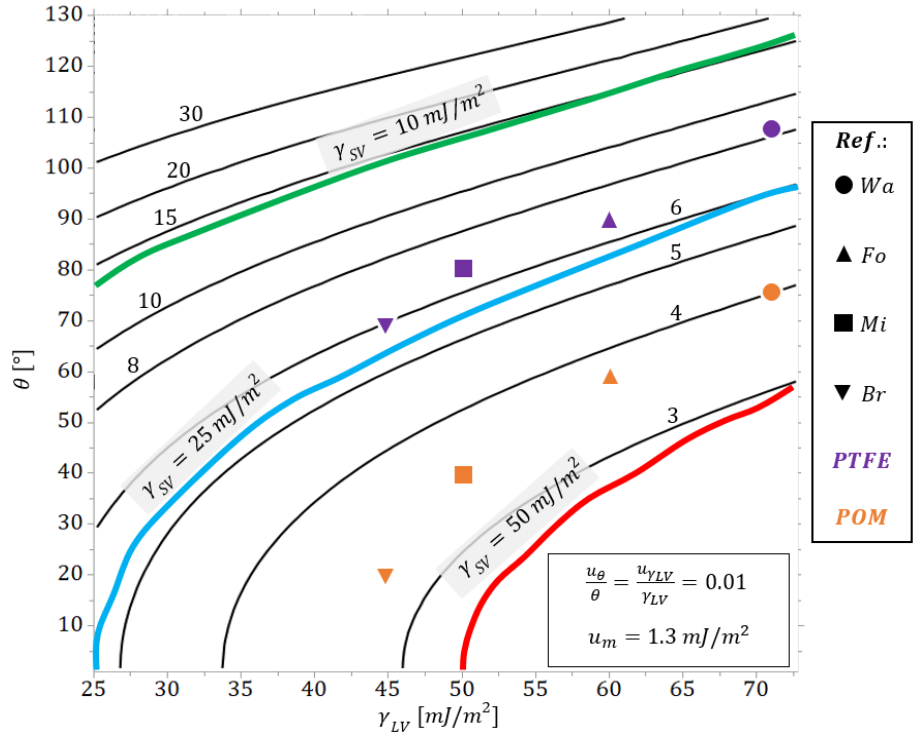

**Figure 10.** *Contour plot of the percentage relative standard uncertainty (equation 29, solid black line) of the surface free energy* $\left(\varepsilon_{u'_{\gamma_{SV}}} = \left(u'_{\gamma_{SV}}/PM\gamma_{SV}\right) \times 100\ \%\right)$ *as a function of* $\theta$ *and* $\gamma_{LV}$. *The plot was built for relative standard uncertainties of* $\theta$ *and* $\gamma_{LV}$ *equal to 1 % each and the systematic standard uncertainty in the Neumann model* $(u_m)$ *of 1.30 mJ/m². The colored lines indicate a given value of* $\gamma_{SV}$ *(green:* $\gamma_{SV}$=10 *mJ/m², light blue:* $\gamma_{SV}$=25 *mJ/m² and red:* $\gamma_{SV}$=50 *mJ/m²).*

## 4. Conclusions

In the first part of this work, we performed a general analysis of the uncertainties of the Neumann EQS (N-III) using the TSM, and obtained the following conclusions:

-The relative standard uncertainty in $\gamma_{SV}$, assuming that the relative uncertainties in $\theta$ and $\gamma_{LV}$ are equal, increases with the value of $\theta$ and is almost independent of the value of $\gamma_{LV}$ of the probe liquid. As a result, from an experimental standpoint, it is possible to reduce the relative uncertainty in $\gamma_{SV}$, for a given solid under study ($\gamma_{SV} = cte$), by using probe liquids with the lowest possible value of $\gamma_{LV}$.

-The analysis of the contact angle uncertainty amplification factor ($UMF_\theta$) allowed concluding that the influence of the relative uncertainty of $\theta$ is magnified when $\theta$ is greater than ~60° (the magnification effect intensifies as the value of $\theta$ increases) and attenuates when it is less than ~60°. This happens for all values of $\gamma_{LV}$. On the other hand, the UMF of the surface tension of the probe liquid $\left(UMF_{\gamma_{LV}}\right)$ shows that the influence of the relative uncertainty of $\gamma_{LV}$ is magnified in practically all the range analyzed and that the magnification effect increases as the values of $\theta$ and $\gamma_{LV}$ increase.

-Assuming that the relative uncertainties in $\theta$ and $\gamma_{LV}$ are equal, the contribution of contact angle uncertainties ($UPC_\theta$) is minimal at low angles and increases with the value of $\theta$ becoming dominant (greater than 50%) when $\theta$ is ~60°, for the whole range of values of $\gamma_{LV}$. On the other hand, the surface tension of the probe liquid $\left(UPC_{\gamma_{LV}}\right)$ shows a behavior opposite to that of $UPC_\theta$.

In the second part of the work, we calculated $\gamma_{SV}$ by means of the N-III EQS on PTFE and POM surfaces, by using four different probe liquids and performing the corresponding propagation of uncertainties of $\theta$ and $\gamma_{LV}$, and obtained the following conclusions:

-The uncertainty of $\theta$ and $\gamma_{LV}$ propagated in $\gamma_{SV}$ is not negligible, generally showing values of ~2 $mJ/m^2$ for the uncertainty expanded to 99 %.

-Some values of $\gamma_{SV}$ obtained from the different probe liquids on the same surface (PTFE or POM) are significantly different even considering the uncertainty in $\theta$ and $\gamma_{LV}$, which allows us to assume the existence of a systematic uncertainty associated with the N–III EQS.

-If the systematic uncertainty associated with the N-III EQS is considered to be equal to 1.3 mJ/m², the differences between the values of $\gamma_{SV}$ obtained with the four different liquids on the same surface are not significantly different, and therefore the model is self-consistent.

-For the cases of PTFE and POM, the consideration of the systematic uncertainty associated with the N-III EQS (equal to 1.3 mJ/m²) has a great impact on the final uncertainty of $\gamma_{SV}$, because it doubles the uncertainty value calculated considering only the uncertainties in $\theta$ and $\gamma_{LV}$.

-Finally, the general uncertainty map proposed and shown in Figure 10 allows the *a priori* determination of the minimum relative uncertainty of $\gamma_{SV}$ for any probe liquid and contact angle.

**Appendix A**

Table A.1 shows the values of surface free energy (with standard and expanded uncertainties) calculated for PTFE and POM, considering the uncertainty associated with the contact angle and surface tension of the probe liquids.

***Table A.1. Surface free energy and standard and expanded uncertainty calculated with different liquids on PTFE and POM.***

| Liquid | Parameters | PTFE | POM |
|--------|------------|------|-----|
| Wa | $\gamma_{SV}$ | 18.23 | 37.33 |
| | $u_{\gamma_{SV}}$ | 0.67 | 0.89 |
| | $\varepsilon_{u_{\gamma_{SV}}}$ | 3.65 | 2.39 |
| | $U_{\gamma_{SV}}$ | 1.72 | 2.30 |
| | $\varepsilon_{U_{\gamma_{SV}}}$ | 9.41 | 6.16 |
| Fo | $\gamma_{SV}$ | 20.11 | 38.02 |
| | $u_{\gamma_{SV}}$ | 0.72 | 0.76 |
| | $\varepsilon_{u_{\gamma_{SV}}}$ | 3.58 | 2.00 |
| | $U_{\gamma_{SV}}$ | 1.86 | 1.97 |
| | $\varepsilon_{U_{\gamma_{SV}}}$ | 9.23 | 5.17 |
| Mi | $\gamma_{SV}$ | 20.37 | 40.33 |
| | $u_{\gamma_{SV}}$ | 0.60 | 0.78 |
| | $\varepsilon_{u_{\gamma_{SV}}}$ | 2.97 | 1.93 |
| | $U_{\gamma_{SV}}$ | 1.56 | 2.01 |
| | $\varepsilon_{U_{\gamma_{SV}}}$ | 7.66 | 4.99 |
| Br | $\gamma_{SV}$ | 23.26 | 42.16 |
| | $u_{\gamma_{SV}}$ | 0.75 | 0.79 |
| | $\varepsilon_{u_{\gamma_{SV}}}$ | 3.23 | 1.88 |
| | $U_{\gamma_{SV}}$ | 1.94 | 2.04 |
| | $\varepsilon_{U_{\gamma_{SV}}}$ | 8.32 | 4.84 |

$\gamma_{SV}$ [mJ/m$^2$]: surface free energy.

$u_{\gamma_{SV}}$ [mJ/m$^2$]: combined standard uncertainty of $\gamma_{SV}$ (equation 8).

$\varepsilon_{u_{\gamma_{SV}}} = \left(u_{\gamma_{SV}}/\gamma_{SV}\right) \times 100$ [%] : relative standard uncertainty of $\gamma_{SV}$.

$k_{\gamma_{SV}}[Adim] = 2.58$ (coverage factor of $\gamma_{SV}$).

$U_{\gamma_{SV}}[mJ/m^2]$: expanded uncertainty of $\gamma_{SV}$ with a confidence level of 99%  (equation 16).

$\varepsilon_{U_{\gamma_{SV}}} = \left(U_{\gamma_{SV}}/\gamma_{SV}\right) \times 100$ [%]: relative expanded uncertainty of $\gamma_{SV}$.

**Appendix B**

Table B.1 shows the values of the difference between the determinations of $\gamma_{SV}$ of the same surface calculated with different liquids ($r_{i-j}$ or $s_{i-j}$). This table also shows the standard uncertainty, the expanded uncertainty for each method and the ratio between the uncertainties obtained by both methods.

***Table B.1. Values of $r_{i-j}$ and $s_{i-j}$, with standard uncertainty, effective degrees of freedom, coverage factors, and expanded uncertainty.***

| Material | Liquids ($i$-$j$) | $r_{i-j}$ or $s_{i-j}$ | Method A | | Method B | | $U_r/U_s$* |
|---|---|---|---|---|---|---|---|
| | | | $u_{r_{i-j}}$ | $U_r$ | $u_{s_{i-j}}$ | $U_s$ | |
| POM | Wa-Fo | -0.68 | 1.06 | 2.75 | 1.17 | 3.03 | 0.91 |
| | Wa-Mi | -2.99 | 1.09 | 2.82 | 1.18 | 3.06 | 0.92 |
| | Wa-Br | -4.82 | 1.13 | 2.91 | 1.19 | 3.07 | 0.95 |
| | Fo-Mi | -2.31 | 0.99 | 2.56 | 1.09 | 2.81 | 0.91 |
| | Fo-Br | -4.14 | 1.07 | 2.75 | 1.10 | 2.83 | 0.97 |
| | Mi-Br | -1.83 | 1.11 | 2.86 | 1.11 | 2.87 | 1.00 |
| PTFE | Wa-Fo | -1.89 | 0.81 | 2.10 | 0.99 | 2.56 | 0.82 |
| | Wa-Mi | -2.14 | 0.71 | 1.84 | 0.91 | 2.36 | 0.78 |
| | Wa-Br | -5.03 | 0.85 | 2.18 | 1.02 | 2.62 | 0.83 |
| | Fo-Mi | -0.25 | 0.77 | 1.97 | 0.94 | 2.42 | 0.81 |
| | Fo-Br | -3.14 | 0.97 | 2.50 | 1.04 | 2.68 | 0.93 |
| | Mi-Br | -2.89 | 0.88 | 2.57 | 0.96 | 2.49 | 0.91 |

$r_{i-j}$ and $s_{i-j}$ $[mJ/m^2]$:  difference between the surface free energy values calculated with methods A and B respectively (equations 19 and 21).

$u_{r_{i-j}}$ and $u_{s_{i-j}}[mJ/m^2]$: standard uncertainty of $r_{i-j}$ and $s_{i-j}$ (equations 20 and 22).

$U_r$ and $U_s$ $[mJ/m^2]$: expanded uncertainty of $r_{i-j}$ and $s_{i-j}$ with a confidence level of 99%. Coverage factor (k) of $r_{i-j}$ and $s_{i-j}$ is 2.58.

* is equal to $u_r/u_s$

**Acknowledgements**

The authors would like to thank the Consejo Nacional de Investigaciones Científicas y Técnicas (CONICET) and Agencia Nacional de Promoción Científica y Tecnológica (ANPCyT) from Argentina. This work was funded by the ANPCyT through PICT-2017-2494 and PUE-CONICET-2019-574-APN projects.